\documentclass[preprint,prd,amsmath,amssymb,amsthm,nofootinbib]{revtex4}
\usepackage[mathscr]{euscript}
\usepackage{textcomp}
\usepackage{bm}
\usepackage{graphicx}
\usepackage{subfigure}
\usepackage{appendix}
\usepackage{multirow}
\usepackage{floatrow}
\usepackage{array}
\usepackage{color}
\usepackage[colorlinks=true,linkcolor=red,urlcolor=blue]{hyperref}
\newcommand{\bea}{\begin{eqnarray}}
\newcommand{\eea}{\end{eqnarray}}
\newcommand{\beq}{\begin{equation}}
\newcommand{\eeq}{\end{equation}}

\def\/{\over}

\begin{document}

\title{Primordial black holes and oscillating  gravitational waves  in slow-roll and slow-climb inflation with an intermediate non-inflationary phase
}

\author{Chengjie Fu}
%
\author{Puxun Wu}
\email{pxwu@hunnu.edu.cn}
\author{Hongwei Yu}
\email{hwyu@hunnu.edu.cn}

\affiliation{Department of Physics and Synergetic Innovation Center for Quantum Effect and
Applications, Hunan Normal University, Changsha, Hunan 410081, People's Republic of China}

\date{\today}

\begin{abstract}

We propose a new single field inflation model  in which the usual slow-roll inflation is joined to a new period of slow-climb and slow-roll inflation through a short intermediate non-inflationary phase.   We then show that  primordial  curvature perturbations can be enhanced at small scales, a sizable amount of primordial black holes (PBHs) can be produced  which  make up most of dark matter,
the gravitational waves (GWs) induced by  scalar metric perturbations that accompany  with the formation of PBHs can be detectable by future GW experiments, and last but not  least, our model is
 compatible with  the latest cosmic microwave background observations.  Remarkably,  the GW spectrum displays a unique  oscillating  character in the ultraviolet regions which originates from the short non-inflationary phase.   A detection of such oscillations in the  GW spectrum may suggest the existence of such a  non-inflationary phase in the whole inflation,  thus providing us a chance to reveal an interesting period in  the evolution of the early Universe and distinguish our model from others.

 \end{abstract}


\maketitle

\section{Introduction.}
Black holes (BHs), whose merger and shadow have been, respectively, listened by LIGO-Virgo observation \cite{PRL116-061102,PRL116-241103} 
  and seen through Event Horizon Telescope \cite{APJ875-L1,APJ875-L2}, are among the most remarkable and fascinating objects in nature. They can be generated from the stellar evolution with masses around $10 M_\odot$,   as well as  from the collapse of over-dense regions induced by large enough curvature perturbations generated during inflation, long before ordinary stellar BHs were  formed. The latter kind of the BHs is dubbed primordial black holes (PBHs)~\cite{Hawking1971,Carr1974,Carr1975,Khlopov2010} and their masses  can span many orders of magnitude contingent upon  the formation mechanism.

PBHs have been proposed as a promising candidate for the origin of BHs detected by the LIGO-Virgo Collaboration \cite{Bird2016,Clesse2017,Sasaki2016,Carr2016} since  many observed binary BHs have masses around $30M_\odot$ which are too heavy for BHs formed by the stellar evolution. For PBHs to be capable of accounting for the LIGO-Virgo gravitational wave (GW) events,  it is required that  they comprise ${\mathcal{O}(1)}$\textperthousand~of the total dark matter  (DM)~\cite{Sasaki2016, Vaskonen}. Actually,  PHBs can make up all DM if their masses are  around  $10^{-16}-10^{-11}M_\odot$ 
~\cite{Katz2018,HSC,Femto,WD,Montero-Camacho:2019jte}. Furthermore, PBHs with a mass around $\mathcal{O}(10^{-5})M_\odot$ can be used to explain  the ultrashort-timescale microlensing events in the OGLE data \cite{OGLE1,OGLE2}.
Recently,  the clustering of orbits of Kuiper belt objects has given some hints that there is a novel object (``Planet 9") with a mass $5-10 M_{\oplus}$ in the outer Solar System~\cite{Nature507-471,AJ151-22,PR01-09}. It has been  suggested that Planet 9 might be a PBH~\cite{1909.11090, Witten}.    As PBHs seem to hold the key to explaining the above mentioned astrophysical and cosmological observations, they are becoming increasingly attractive in physics, astronomy and cosmology.

The production of a sizable amount of PBHs requires that the power spectrum of primordial  curvature perturbations $\mathcal{P_R}$ produced during the inflationary era is amplified about seven orders of magnitude compared with  that,  $\mathcal{P_R}\sim \mathcal{O}(10^{-9})$, on the cosmic microwave background (CMB) scales. Since in the slow-roll inflation model the power spectrum of primordial curvature perturbations is inversely proportional to one of the slow-roll parameters, a natural mechanism to enhance the curvature perturbations is to flatten the potential to realize a period of  ultra-slow-roll inflation, i.e., the so-called inflection-point inflation~\cite{Bellido2017,Germani2017,Hu2017,Ezquiaga2018,Gong2018, Ballesteros2018, Guo2020, Drees2019, Dalianis2019, Kannike,Rasanen:2018fom}.  The ultra-slow-roll inflation can also be obtained through slowing down the motion of the inflaton by increasing friction~\cite{Fu2019} or introducing a non-canonical kinetic term~\cite{Lin2020}.  In addition, it has been found that the primordial curvature perturbations can also be enhanced through some other mechanisms, such as making  the sound speed  to approach to zero during some stages of inflation~\cite{Ballesteros2019, Kamenshchik2019}, the parametric resonance arising from the oscillating sound speed squared~\cite{Cai2018,Chen:2019zza,Chen:2020uhe} or the oscillating potential~\cite{Cai2020}, and so on.  Associated with the formation of PBHs, the enhanced curvature perturbations, when they reenter the horizon,  may also induce very large scalar metric fluctuations, which would generate a significant stochastic GW background \cite{Kohri2018, Inomata2019, RG2019_1, RG2019_2, Bartolo:2018evs, Bartolo:2018rku, YF2019, Wang2019, Lu2019,Yuan2019,Fu2020, Unal, Unal2, Braglia, Saito09,Saito10}. These scalar induced GWs (SIGWs) are very important since they might be detected by future GW projects.

In the standard inflation theory,  the cosmic inflation is followed by a so-called reheating phase, in which  the inflaton rolls rapidly through  the  minimum of its potential and oscillates around it to thermalize the Universe. As we have already mentioned, one way to amplify the curvature perturbations at small scales so that a sizable amount of PBHs  will be generated is to slow down the rolling of the inflaton during the late stage of inflation either by flattening the potential or increasing the friction. Here, we propose a new inflation model  which also naturally achieves the goal of enhancing curvature perturbations at small scales  and at the same time satisfies the latest CMB observational constraints. Our basic idea is that the inflaton,
after the usual slow-roll phase and passing across the  minimum of its potential, undergoes a new period of slow-climb and slow-roll  as it climbs up a flattened potential, which is then followed by the standard reheating.  In our model, the usual slow-roll inflation is joined to a
new slow-climb and slow-roll inflation with a short non-inflationary phase in between, i.e, altogether, there exist three inflationary phases and one non-inflationary phase in the whole inflation era (see  Fig.~\ref{fig1} for a graphic description).  Since the slow-climb phase is a non-attractor inflation due to the  rapid change of the velocity of the inflaton,  the curvature perturbations can be enhanced so as to produce a sizable amount of the PBHs.  In the following, we construct  a concrete simple single field model and study its consequences. As we will show later, the generated PBHs can make up most of DM and the SIGWs can be detectable by future GW projects. Remarkably, as a result of the short non-inflationary phase in the whole inflation era, the spectrum of the SIGWs displays an oscillating behavior  in the ultraviolet regions which is testable by future GW experiments.

\section{Model}
\begin{figure}
\centering
\includegraphics[width=0.8\textwidth ]{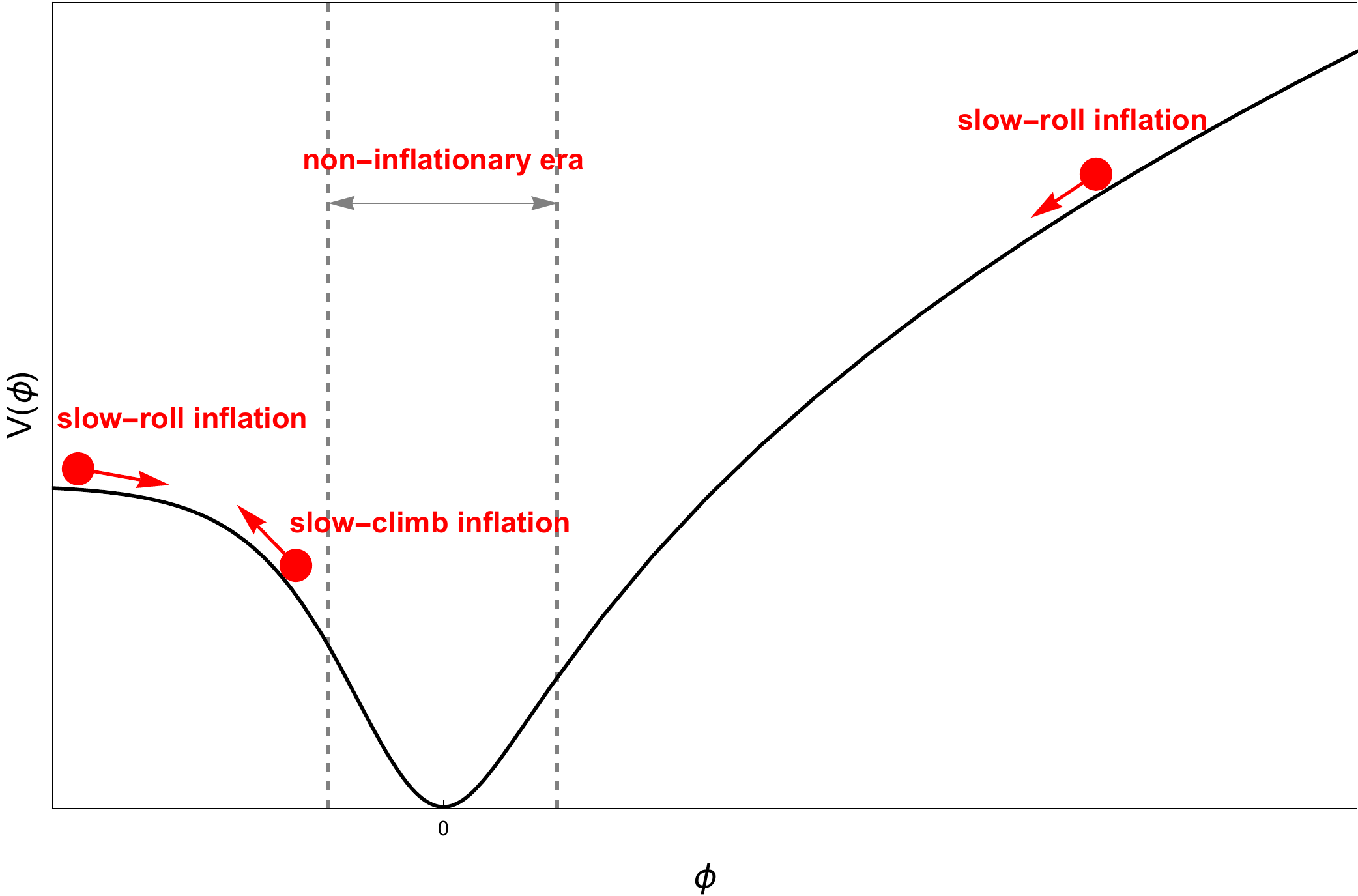}
\caption{\label{fig1} The  diagram of potential. }
\end{figure}

To realize a non-attractor inflation to enhance the curvature perturbations after the inflaton rolls through the minimum of its potential for the first time, we consider the following potential
\bea\label{potential}
V(\phi) &=& \Lambda^4 \left\{ \left[ \left(1+\frac{\phi^2}{M_1^2} \right)^{\frac{q}{2}} -1 \right]\Theta(\phi) + \xi \tanh^2\left( \frac{\phi}{M_2}\right)\Theta(-\phi) \right\}
\eea
with $q=2/5$. Here, $\Lambda$, $M_1$ and $M_2$ are the parameters with the dimension of mass, $\xi$ is a dimensionless parameter, and $\Theta$ is the Heaviside theta function. Let us note that the  $\phi>0$  part of this potential (\ref{potential})  is of the monodromy type \cite{Silverstein2008,McAllister2014} while  the other part is similar to that of an $\alpha$-attractor T-model~\cite{Kallosh2013}.  To ensure that the potential (\ref{potential}) has a continuous second derivative with respect to $\phi$ at $\phi=0$, parameter $\xi$ is determined to be $\xi = q(M_2/M_1)^2/2$. So,  parameters $M_1$ and $M_2$ control the inflationary dynamics. The potential is plotted in Fig.~\ref{fig1}.

A successful inflation must satisfy the CMB observations. The Planck results \cite{1Planck2018} have constrained  the amplitude of the power spectrum, the scalar spectral index and the tensor-to-scalar ratio at the pivot scale $k_\ast=0.05\;\mathrm{Mpc}^{-1}$ to be $ \ln{(10^{10}\mathcal{P}_\mathcal{R})}=3.044\pm0.014 ~(68\%\;\mathrm{C.L.}),n_s=0.9649\pm0.0042 ~(68\%\;\mathrm{C.L.}),$ and$ r<0.07 ~(95\%\;\mathrm{C.L.})$, respectively.
 To obtain the allowed region of  $M_1$ (or $M_2$) constrained by   $n_s$, we need to give  a concrete value to $M_2$ (or $M_1$). Thus,  there is only one free parameter in our model since $\Lambda$ is determined only by $\mathcal{P_R}$ after  $M_1$ and $M_2$ are given. Since parameter $M_2$ is decisive to enhancing the curvature perturbations, we  take  $M_2$ as the free parameter. In our analysis, we choose $M_2=0.009M_{\mathrm{p}}$, where $M_{\mathrm{p}}$ is the reduced Planck mass and $M_{\mathrm{p}}=1$ is set in the numerical calculations. 
After setting the \textit{e}-folding number from the time when $k_\ast$ exits the horizon to the end of the whole inflation as $N_\ast=60$, we find that $M_1$ is constrained to be $0.00179670< M_1/M_{\mathrm{p}} <0.00179694$ by $n_s$. We adopt $M_1 = 0.00179690 M_{\mathrm{p}}$ as an example in the our analysis. In this case, we find that  $n_s\simeq0.9625$ and $r\simeq0.051$, which  are compatible with the Planck CMB observations~\cite{1Planck2018}.   Using $\mathcal{P_R}\simeq 2.1\times 10^{-9}$ at the scale $k_\ast$, we obtain $\Lambda \simeq 0.00286$.

\begin{figure}
\centering
\subfigure{\label{fig2a}}{\includegraphics[width=1.01\textwidth ]{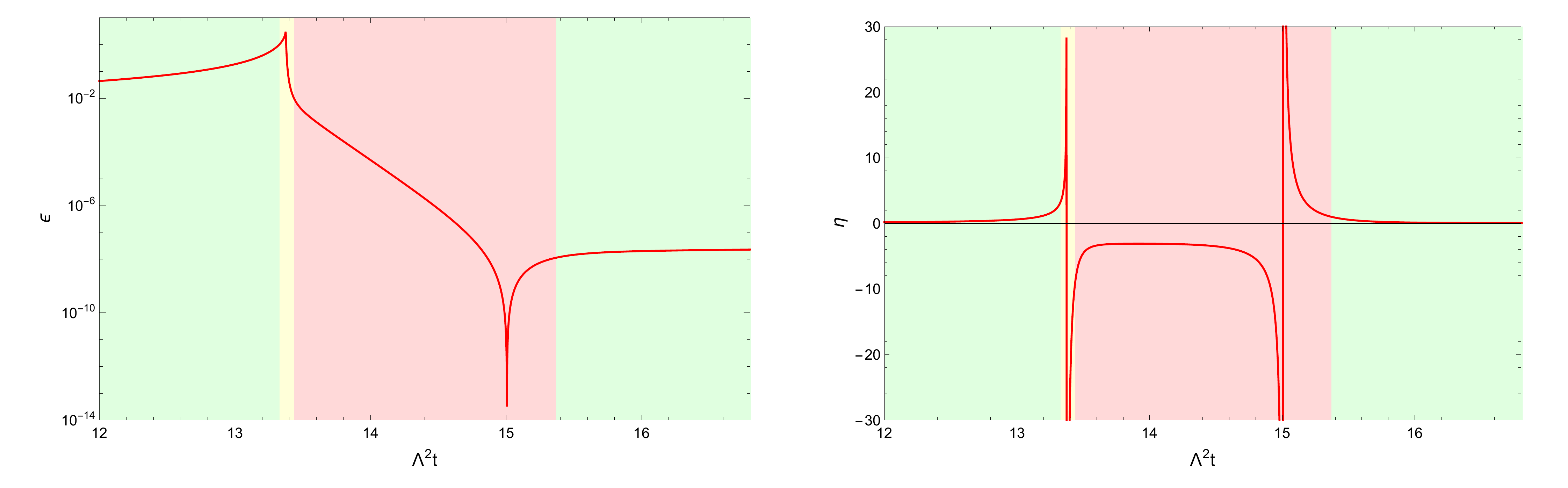}}
\caption{\label{fig2} The evolutions of the slow-roll parameters $\epsilon$ and $\eta$ as a function of the rescaled time $\Lambda^{2}t$ (red line). The light-green and light-red regions represent  the slow-roll inflationary phase and the non-attractor one, respectively. The light-yellow region shows the non-inflationary stage.}
\end{figure}

Now, we discuss  the inflationary dynamics of our model. Initially, the inflaton rolls down its potential from $\phi>0$. We set $t_i=0$ as the time when $k_\ast$ exits the horizon. Figure \ref{fig2} shows the evolutions of slow-roll parameters $\epsilon=-\dot H/H^2$ and $\eta=\ddot\phi/(H\dot\phi)$ with respect to the rescaled time $\Lambda^2t$, where   $H=\frac{\dot{a}}{a}$ with $a$ being the scale factor and an overdot denoting the derivative with respect to the cosmic time $t$. The whole evolution can be divided into four stages, which are described  in Figs.~\ref{fig1} and~\ref{fig2} with four different color regions.  The first stage is the usual slow-roll inflation in which the inflaton rolls  slowly down the monodromy type potential.  This stage ends at the time $t=t_1\simeq13.328\Lambda^{-2}$ ($\epsilon=1$ when $t=t_1$) and generates approximately $\triangle N \simeq 33.224$ \textit{e}-folds for model parameters given the above. Then,  the inflaton rolls rapidly through the minimum of its potential and  climbs up the  potential. The velocity of the inflaton decreases quickly during the climb-up.  When the slow-roll parameter $\epsilon$ is very small, a new inflation stage begins.   Taking the time $t_2\simeq13.436\Lambda^{-2}$ at which $\epsilon=0.01$ as the onset time of  this  new slow-climb inflation, the Universe undergoes a short non-inflationary phase in the period of $t_1<t<t_2$,  generating just $\triangle N \simeq 0.144$ \textit{e}-folds. The new slow-climb inflation which follows the non-inflationary phase is a non-attractor inflation since the velocity of the inflaton changes  rapidly, which can be seen from the right panel of Fig.~\ref{fig2}. This non-attractor inflation goes on until $t=t_3\simeq15.368\Lambda^{-2}$ ($\eta=1$ when $t=t_3$), generating about $\triangle N \simeq 2.499$ \textit{e}-folds. The last stage ($t>t_3$) is a period of the usual slow-roll inflation again, which generates about $\triangle N \simeq 24.133$ \textit{e}-folds. After this stage, the Universe enters the standard reheating phase in which the inflaton oscillates rapidly around the minimum of the potential to thermalize the Universe. Let us note here that the simplest potential that can lead to an inflation era similar to ours is  the double-well potential of chaotic inflation when fine-tuned so that the inflaton, after passing through the first minimum, can slowly climb over the top of the Mexican hat and roll down to the second minimum generating new inflationary phases,   and  oscillate around the second minimum  to reheat  the Universe,  similar to what was proposed in~\cite{Yokoyama98, Yokoyama99, Yokoyama08}. Unfortunately, this simplest scenario of chaotic inflation is ruled out by the latest CMB observation constraints on the scalar spectral index and the tensor-to-scalar ratio. Moreover, even the CMB constraints  are put aside,  one can show that to produce a sizable amount of  PBHs that can explain the most dark matter  the parameters of the potential have to be  fine-tuned. For example, parameter
$\nu$ has to be tuned to $10^{-6} $.  By contrast, a precision up to only  $10^{-3}$ is needed  for our free parameter $M_2$.  In addition, some double inflation models have also been proposed to produce PBHs~\cite{Kannike, Inomata, Yamaguchi}. 

To produce PBHs, the curvature perturbations at the small scales must be enhanced during inflation. The curvature perturbation $\mathcal{R}_k$ in the momentum space satisfies the equation of motion \bea\label{A1}
\ddot{\mathcal{ R}}_k +\left(3+ 2\epsilon+2\eta\right)H\dot{\mathcal{ R}}_k+ \frac{k^2}{a^2}\mathcal{R}_k = 0\;.
\eea
 During the usual slow-roll inflation, the third term can be neglected  at the super Hubble horizon scales. So,  the general solution of this equation contains a constant term and a decaying term, which results in  a nearly scale-invariant power spectrum. While in the non-attractor inflation phase since the coefficient of the second term in Eq.~(\ref{A1}) becomes  negative,  the curvature perturbations will be increased. For the modes which have left the horizon before  several   \textit{e}-folds at the end of the first slow-roll inflation, their growth during the non-attractor inflation can not  overtake the decay, and thus they remain nearly constant, which gives a scale-invariant power spectrum at the large scales to satisfy the CMB observations.  For the modes leaving the horizon during the last several \textit{e}-folds of the first slow-roll inflation and  the slow-climb inflation,  the non-attractor inflation will make them grow significantly.  So, the power spectrum at these scales will be enhanced.

\begin{figure}
\centering
\includegraphics[width=1.01 \textwidth ]{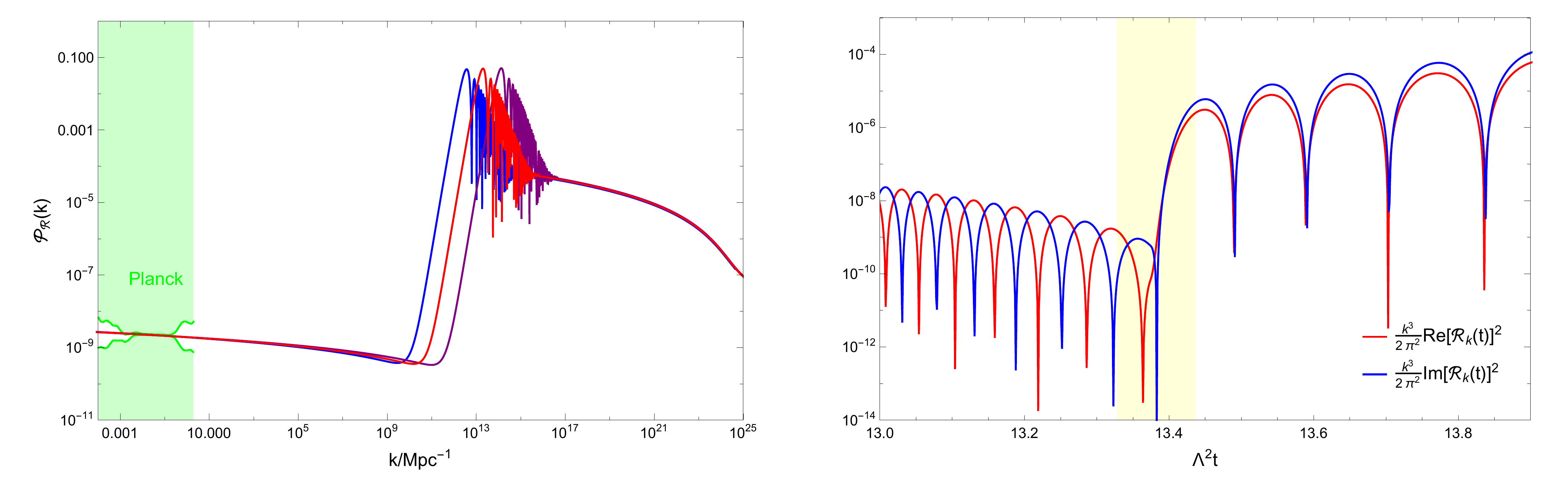}
\caption{\label{fig3}Left:  The resulting power spectrum $\mathcal{P_R}$ (red line). For a comparison, the cases of $M_2=0.0085M_{\mathrm{p}}$ with $M_1=0.001665856M_{\mathrm{p}}$ and $M_2=0.0095M_{\mathrm{p}}$ with $M_1=0.00193068M_{\mathrm{p}}$ are shown with the purple and blue curves, respectively.
The green-shaded region is excluded by the current CMB observations \cite{1Planck2018}.  Right: The time evolutions of $k^3\mathrm{Re}[\mathcal{R}_k(t)]^2/(2\pi^2)$ (red line) and $k^3\mathrm{Im}[\mathcal{R}_k(t)]^2/(2\pi^2)$ (blue line) for $k=2\times10^{14}\mathrm{Mpc}^{-1}$. $\mathrm{Re}[\mathcal{R}_k(t)]$ and $\mathrm{Im}[\mathcal{R}_k(t)]$ denote the real part and imaginary part of $\mathcal{R}_k(t)$, respectively. The light-yellow region represents the non-inflationary stage. }
\end{figure}

The left panel of Fig.~\ref{fig3} shows the resulting power spectrum $\mathcal{P_R}$. One can see that the power spectrum remains nearly scale invariant on the large scales, which is compatible with the Planck 2018 results. At the small scales, the power spectrum is enhanced by the non-attractor inflation. 
The enhanced power spectrum displays an oscillating behavior, which is a result of the existence  of the non-inflationary stage. During this stage, the second term in Eq.~(\ref{A1})  becomes dominant and its coefficient turns negative in the latter part of this stage since $\eta$ is much less than $0$, which results in some modes to grow in a short time interval. As a consequence, the phase difference between the real and imaginary parts of the modes whose wave lengths are in the horizon but not too small changes from $\pi/2$ to near $0$ after they undergo this short non-inflationary phase.  This can be seen from the right panel of Fig.~\ref{fig3}. Accordingly, $|\mathcal{R}_k(t)|^2$ will evolve into an oscillating era, which leads to the oscillating structure in the power spectrum. This is a unique character for the model proposed in this paper.  In order to check the effect of a small change of $M_2$ on our result, we perform a comparison with the cases of $M_2=0.0085M_{\mathrm{p}}$ with $M_1=0.001665856M_{\mathrm{p}}$ and $M_2=0.0095M_{\mathrm{p}}$ with $M_1=0.00193068M_{\mathrm{p}}$ in the left panel of Fig.~\ref{fig3}. Obviously, these power spectra exhibit the similar shape and peak values, which means that small variations of $M_2$ do not result in significant difference in the power spectrum.

\section{Primordial black holes}
When the enhanced curvature perturbations  reenter the horizon, they may cause  over-dense regions to collapse into BHs. PBHs have been proposed as a component of DM. The current fraction of PBHs against the total DM can be expressed as
\bea\frac{\Omega_{\mathrm{PBH}}}{\Omega_{\mathrm{DM}}}= \int \frac{dM}{M} f_{\mathrm{PBH}}(M)\;, \eea
where $M$ is the PBH mass, 
 which is relative to  the horizon mass at the horizon entry of the perturbations through  the relation $M(k)= \gamma \frac{4\pi }{\kappa^2H} \big|_{k=aH} \simeq M_\odot\left(\frac{\gamma}{0.2}\right)\big(\frac{g_\ast}{10.75}\big)^{-\frac{1}{6}} \big(\frac{k}{1.9\times10^6\;\mathrm{Mpc}^{-1}} \big)^{-2}\;$,  
 and 
 \bea\label{f_pbh} f_{\mathrm{PBH}}(M)&\equiv& \frac{1}{\Omega_{\mathrm{DM}}} \frac{d\Omega_{\mathrm{PBH}}}{d\ln{M}} \simeq \frac{\beta(M)}{1.84\times10^{-8}}\left(\frac{\gamma}{0.2}\right)^{\frac{3}{2}} \left(\frac{10.75}{g_\ast}\right)^{\frac{1}{4}}\left(\frac{0.12}{\Omega_{\mathrm{DM}}h^2}\right)\left(\frac{M}{M_\odot}\right)^{-\frac{1}{2}}\;. \eea
Here, $\Omega_{\mathrm{DM}}$ is the current density parameter of DM, which is limited to be  $\Omega_{\mathrm{DM}}h^2\simeq0.12$ by the Planck 2018 observations \cite{2Planck2018} and  $g_\ast$  is the number of effectively relativistic degrees of freedom at the PBH formation. Assuming that the PBHs are formed deep in the radiation-dominated era, one has  $g_\ast=106.75$. Parameter $\gamma$ is the ratio of the PBH mass to the horizon mass and represents the efficiency of collapse, which,  in our analysis, is set to be  $\gamma\simeq (1/\sqrt{3})^3$ estimated by the simple analytical calculation \cite{Carr1975}.

Function $\beta(M)$ in Eq.~(\ref{f_pbh}) is the production rate of PBHs with mass $M(k)$. Based on the Press-Schechter theory, one has $\beta(M)=\int_{\delta_c}\frac{d\delta}{\sqrt{2\pi\sigma^2(M)}}e^{-\frac{\delta^2}{2\sigma^2(M)}}=\frac{1}{2}\mathrm{erfc}\big(\frac{\delta_c}{\sqrt{2\sigma^2(M)}}\big)\; $ if the probability distribution function of perturbations is assumed to be Gaussian~\cite{Young2014}. Here, erfc denotes the complementary error function and $\delta_c$ is the threshold of the density perturbations for the PBH formation.  We adopt $\delta_c=0.43$  in our subsequent calculations since recent studies \cite{Musco2013,Harada2013} have shown that $\delta_c\simeq0.4 \sim 0.5$.  The variance $\sigma^2(M)$ represents the coarse-grained density contrast with the smoothing scale $k$, which is defined to be $\sigma^2(M(k))= \int d\ln{q} \ W^2(qk^{-1})\frac{16}{81}(qk^{-1})^4\mathcal{P_R}(q)$~\cite{Young2014}, where  $W$ is the window function and is  taken to be  the Gaussian function in our analysis.

\begin{figure}
\centering
\includegraphics[width=1.01\textwidth ]{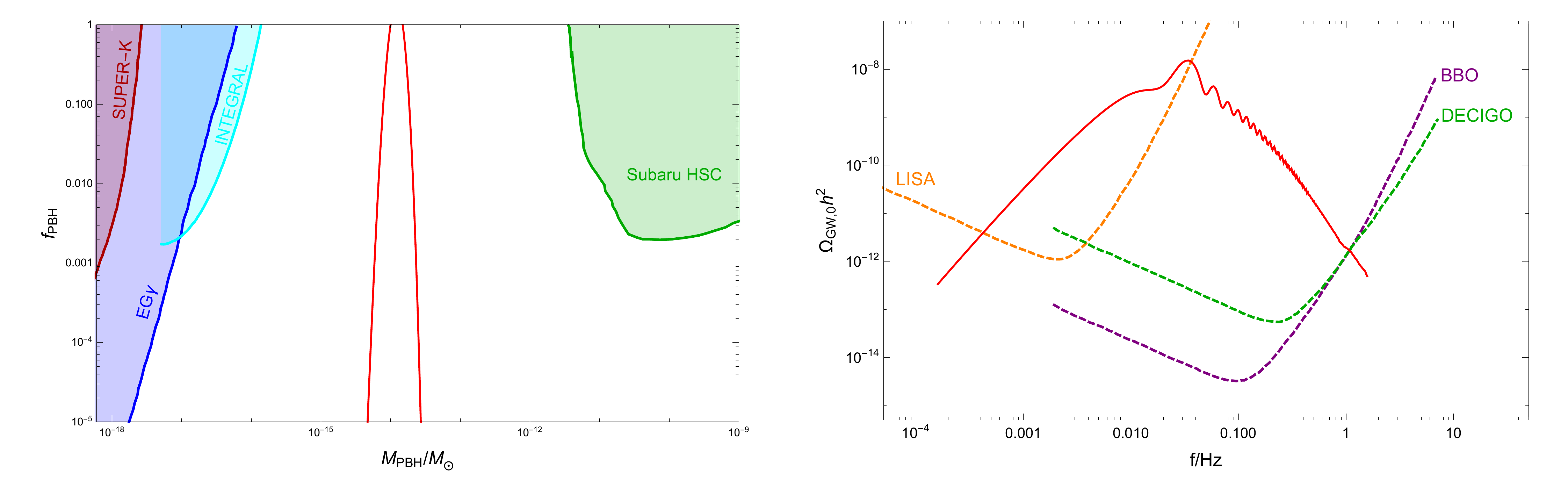}
\caption{\label{fig5} Left: The predicted mass spectrum of PBHs (red line). The shaded regions show the current observational constraints: extragalactic gamma rays from PBH evaporation (EG$\gamma$) \cite{Carr2010}, galactic center 511 keV gamma-ray line (INTEGRAL) \cite{Laha2019}, 
diffuse supernova neutrino background (SUPER-K)~\cite{super-k}, and microlensing events with Subaru HSC (Subaru HSC) \cite{Niikura2019}. Right:  The current energy spectrum of the SIGWs predicted by our model (red line). The dashed lines are the expected sensitivity curves of the future GW projects.}
\end{figure}

The predicted mass spectrum of PBHs, $f_{\mathrm{PBH}}(M)$,  is plotted in the left panel of Fig.~\ref{fig5}, in which the current observational constraints on the PBH abundance are also shown. We find that the  mass spectrum of PBHs has a sharp peak at $1.22\times10^{-14}M_\odot$, and its height is about $1.895$. These PBHs can make up most of DM since $\frac{\Omega_{\mathrm{PBH}}}{\Omega_{\mathrm{DM}}}\simeq 0.84$.



\section{Scalar induced gravitational waves}
When the enhanced curvature perturbations reenter the horizon,  the PBHs will be formed in the over-dense regions, and  at the same time very large scalar metric perturbations will be induced. These large metric perturbations become a significant GW source and emit abundant GW signals via the second-order effect to form a stochastic GW background. It has been found that, during the radiation-dominated era, the energy density of SIGWs per logarithmic interval of $k$ can be evaluated, at conformal time $\tau_c$ which represents the time when $\Omega_{\rm{GW}}$ stops to grow, as \cite{Kohri2018}
\bea\label{OGW}
\Omega_{\rm{GW}}(\tau_c,k) = \frac{1}{12} \int^\infty_0 dv \int^{|1+v|}_{|1-v|}du   I(u,v) 
\mathcal{P}_\mathcal{R}(ku)\mathcal{P}_\mathcal{R}(kv)
\eea
with
\bea I(u,v) &=& \left( \frac{3(u^2+v^2-3)}{4u^3v^3}\right)^2  \left( \frac{4v^2-(1+v^2-u^2)^2}{4uv}\right)^2\nonumber \\ 
&&\bigg\{\left[-4uv+(u^2+v^2-3) \ln\left| \frac{3-(u+v)^2}{3-(u-v)^2}\right| \right]^2 \nonumber \\ && \quad + \pi^2(u^2+v^2-3)^2\Theta(v+u-\sqrt{3})\bigg\}\;.
\eea
The current energy spectra of SIGWs can be expressed as~\cite{Inomata2019}
\bea\label{OGW0}
\Omega_{\rm{GW},0}h^2 = 0.83\left( \frac{g_*}{10.75} \right)^{-1/3}\Omega_{\rm{r},0}h^2\Omega_{\rm{GW}}(\tau_c,k)\;,
\eea
where $\Omega_{\rm{r},0}h^2\simeq 4.2\times 10^{-5}$ is the current density parameter of radiation.  The  current frequency $f$ of SIGWs, which relates to the comoving wave number $k$, can be obtained  through  $
f=1.546\times10^{-15}\frac{k}{1\rm{Mpc}^{-1}}\rm{Hz}\;$.

We show the predicted current energy spectra of SIGWs associated with the production of PBHs in the right panel of Fig. \ref{fig5}. One can see that these GW signals could all be probed by future GW projects, i.e., LISA~\cite{LISA},  DECIGO~\cite{DECIGO}, and BBO~\cite{BBO}. It is interesting to note that the GW power spectrum displays a special oscillating  structure in the ultraviolet regions, which is a unique feature that distinguishes our model from others. Since these oscillations originate from the short non-inflationary stage, if they are found in the future GW experiments, we may conclude that there exists  a non-inflationary phase in the whole inflation era in the early evolution of our Universe.

\section{Conclusion}
In this paper, we propose  a simple single field inflation model, which can enhance naturally the  curvature perturbations so that  a sizable amount of PBHs can be produced. In our model,  the whole inflation  contains four stages: two slow-roll inflation phases, one slow-climb inflation phase, and a short non-inflationary one.  Since the slow-climb phase is  a non-attractor inflation,  the curvature perturbations can be enhanced  during this stage to satisfy the condition for a production of a sizable amount of PBHs. We find that the produced PBHs can make up most of DM, and SIGWs accompanying with the formation of PBHs can be detectable by future GW experiments, i.e. LISA, BBO, and DECIGO. Remarkably, the spectrum of SIGWs shows a unique  oscillating  character in the ultraviolet regions, which is a result of  the short non-inflationary phase  in our model. A detection of such oscillations in the GW spectrum may suggest the existence of such a  non-inflation phase in the whole inflation,  thus providing us a chance to reveal an interesting period in the  evolution of the early Universe and distinguish our model from other inflation models.

\begin{acknowledgments}
This work was supported in part by the NSFC under Grants No. 11775077, No. 11435006, No. 11690034, and No. 11805063, and by the Science and Technology Innovation Plan of Hunan province under Grant No. 2017XK2019.
\end{acknowledgments}

\end{document}